\documentclass[a4paper]{article}

\usepackage{INTERSPEECH2021}
\usepackage{graphicx}
\usepackage{amssymb, amsmath, bm, mathtools,array}
\usepackage{textcomp}
\usepackage{hyperref}
\usepackage{verbatim}
\usepackage{lipsum} 
\usepackage{multirow}
\usepackage{siunitx}
\usepackage{diagbox}
\usepackage{dsfont}
\usepackage{colortbl}
\usepackage{booktabs}
\usepackage{caption}
\usepackage{subcaption}

\newcommand{\bs}[1]{\boldsymbol{#1}}

\RequirePackage{color}\definecolor{BLUE}{rgb}{0,0,0}

\newcommand{\textcirclednew}[1]{\raisebox{.5pt}{\textcircled{\raisebox{-.9pt} {#1}}}}

\newcommand{\absI}{\texttt{abs-mel-noi}}
\newcommand{\absII}{\texttt{abs-mfb-noi}}
\newcommand{\absIII}{\texttt{abs-mel-sin}}
\newcommand{\absIV}{\texttt{abs-mfb-sin}}
\newcommand{\sysI}{\texttt{midi-noi-nsf}}
\newcommand{\sysII}{\texttt{midi-sin-nsf}}
\newcommand{\sysIII}{\texttt{pfnet-mel-noi}}
\newcommand{\sysIV}{\texttt{pfnet-mfb-noi}}
\newcommand{\sysV}{\texttt{pfnet-mel-sin}}
\newcommand{\sysVI}{\texttt{pfnet-mfb-sin}}
\newcommand{\sysVII}{\texttt{taco2-mfb-noi}}
\newcommand{\sysVIII}{\texttt{taco2-mfb-sin}}
\newcommand{\sysIX}{\texttt{taco3-mfb-noi}}
\newcommand{\sysX}{\texttt{taco3-mfb-sin}}
\newcommand{\sysXI}{\texttt{taco4-mfb-noi}}
\newcommand{\sysXII}{\texttt{taco4-mfb-sin}}
\newcommand{\sysXIII}{\texttt{pfnet-spec-GL}}
\newcommand{\sysXIV}{\texttt{Fluidsynth}}
\newcommand{\sysXV}{\texttt{Pianoteq}}

\title{Text-to-Speech Synthesis Techniques for  MIDI-to-Audio Synthesis}
\name{Erica Cooper$^\dagger$\thanks{$^\dagger$Equal contribution}, Xin Wang$^\dagger$\footnotemark[1], Junichi Yamagishi}
\address{National Institute of Informatics, Chiyoda-ku, Tokyo, Japan}
\email{ecooper@nii.ac.jp, wangxin@nii.ac.jp, jyamagis@nii.ac.jp}

\begin{document}

\maketitle
\begin{abstract}
Speech synthesis and music audio generation from symbolic input differ in many aspects but share some similarities. In this study, we investigate how text-to-speech synthesis techniques can be used for piano MIDI-to-audio synthesis tasks. Our investigation includes Tacotron and neural source-filter waveform models as the basic components, with which we build MIDI-to-audio synthesis systems in similar ways to TTS frameworks. We also include reference systems using conventional sound modeling techniques such as sample-based and physical-modeling-based methods. The subjective experimental results demonstrate that the investigated TTS components can be applied to piano MIDI-to-audio synthesis with minor modifications. The results also reveal the performance bottleneck -- while the waveform model can synthesize high quality piano sound given natural acoustic features, the conversion from MIDI to acoustic features is challenging. The full MIDI-to-audio synthesis system is still inferior to the sample-based or physical-modeling-based approaches, but we encourage TTS researchers to test their TTS models for this new task and improve the performance.
\end{abstract}

\section{Introduction}



Speech and music are human universals, and they have been the theme of numerous research topics in the social and natural sciences. From an engineering perspective, 
both speech and music information processing deal with symbolic and acoustic data, i.e., symbolic music notes or text, and acoustic music or speech audio signals. 
This similarity makes it possible to share methodologies across disciplines, especially those based on data-driven deep learning. For example, AI music composition, which learns a distribution of music notes to generate new songs, is based on language models for text data. Automatic audio transcription, which converts music into a sequence of notes, uses similar techniques to speech recognition such as Viterbi decoding and DNNs for classification tasks. 

From the music notes to instrumental audio signals\footnote{In this paper, we focus on musical instrument sounds rather than singing voice synthesis.}, however, cross-disciplinary techniques are less explored even though the concept is very similar to text-to-speech (TTS) synthesis. 
It has not been until recently that some deep learning models, such as WaveNet \cite{oord2016wavenet, hawthorne2018enabling}, have been used in both tasks. While research in both fields has investigated other related models such as GAN \cite{engel2019gansynth, kumar2019melgan} and VAE \cite{roche2018autoencoders, bitton2020vector}, most studies focus on music audio-to-audio mapping tasks. The question of how and to what extent TTS approaches can be applied to music audio generation remains to be explored. 

This study is our initial step to address the aforementioned question. We focus on MIDI-to-audio synthesis for piano because of the available data, but the methodology is expected to be applicable to many other instruments.
In Section~\ref{seq:2}, we compare TTS to music audio generation from MIDI and explain the possibility of using TTS methods for the MIDI synthesis task. In Section~\ref{seq:3}, we explain the MIDI-to-audio systems that use many components from TTS, including Tacotron \cite{Wang2017} and neural source-filter (NSF) waveform model \cite{wangNSFall}. We introduce modifications to those components to account for the intrinsic differences between music and speech. Furthermore, we introduce MIDI-specific acoustic features and compare them with the Mel-spectrogram for MIDI-to-audio synthesis.

Based on a subjective evaluation, our study tentatively suggests that many TTS techniques can be adapted to music audio generation with slight modifications, and we observed trends similar to those in TTS. For audio generation, the NSF models, which were originally created for speech modeling, can produce high-quality polyphonic piano sound in the copy-synthesis scenario.
For acoustic modeling, the Tacotron-based models demonstrated competitive performance to produce acoustic features from the MIDI piano roll input.  
However, the conversion from MIDI to acoustic features is the most challenging task, similar to the bottleneck in TTS. 
Hence, we encourage TTS researchers to extend their research outcomes to the music audio generation task.

\section{Using TTS techniques for MIDI-to-audio synthesis}
\label{seq:2}
\begin{figure}[t]
\centering
{\includegraphics[trim=10 470 10 75, clip, width=0.46\textwidth]{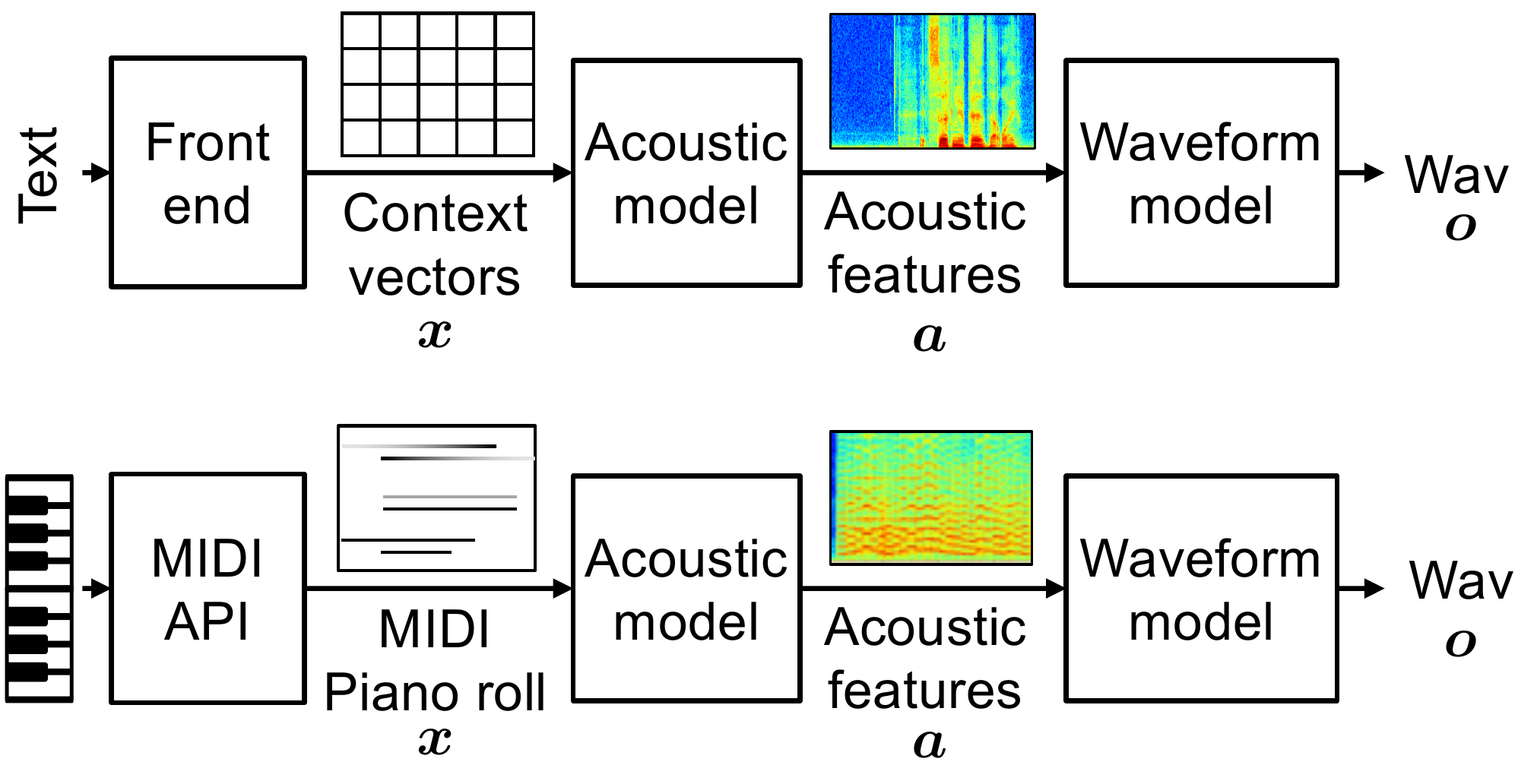}}
\begin{tabular}{rrl}
\multirow{2}{*}{Synthesis process $\Big\{$} & Aligned & $\bs{x}_{1:N} \rightarrow \bs{a}_{1:N} \rightarrow \bs{o}_{1:T}$ \\
& Unaligned & $\bs{x}_{1:M} \rightarrow \bs{a}_{1:N} \rightarrow \bs{o}_{1:T}$ \\
\end{tabular}
\vspace{-3mm}
\caption{Comparing TTS and MIDI-to-audio synthesis. Temporal length of synthesized waveform $\bs{o}_{1:T}$ and acoustic feature $a_{1:N}$ is related by the frame rate $L$, i.e., $T=N\times{L}$.}
\label{fig:app_example}
\end{figure}

%
This paper focuses on music audio generation from music transcription data in MIDI format. As illustrated in Figure~\ref{fig:app_example}, the MIDI-to-audio synthesis process is similar to TTS since both convert symbolic data into audio signals. 

\begin{figure}[t]
\centering
{\includegraphics[trim=10 570 10 80, clip, width=0.46\textwidth]{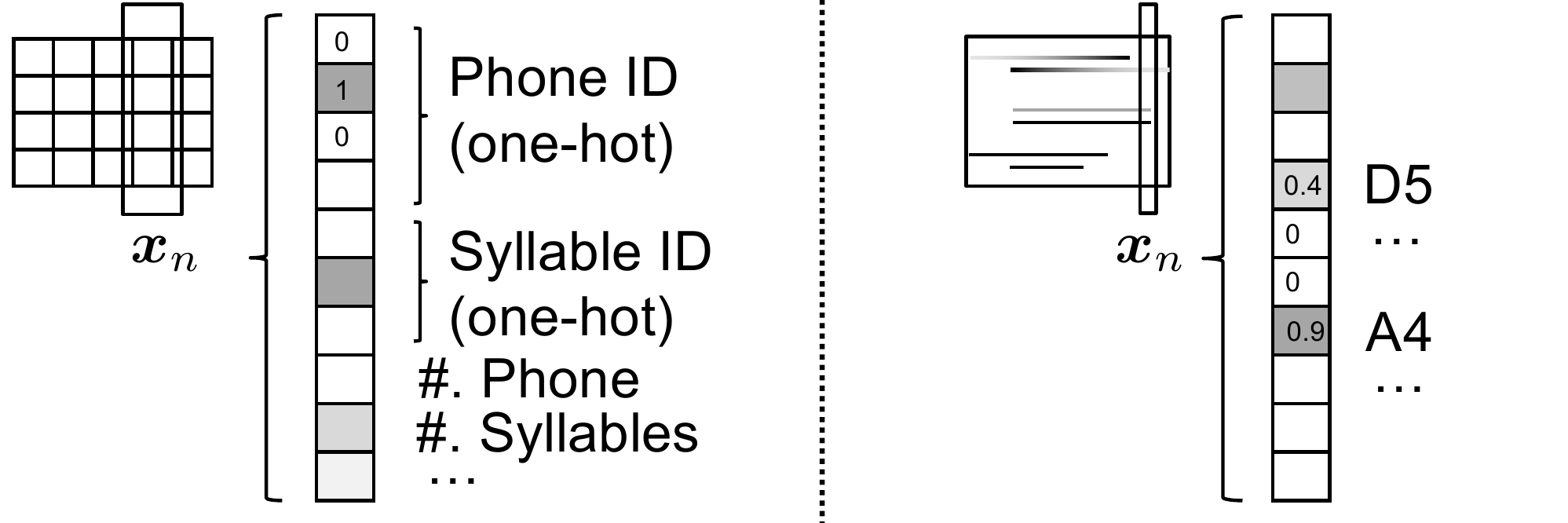}}
\vspace{-3mm}
\caption{Comparing contextual vector for statistical parametric TTS (left) and MIDI piano roll (right).}
\label{fig:input_example}
\vspace{-5mm}
\end{figure}

In both cases, the front-end converts the input text or MIDI raw data into a representation as input to the acoustic model. In the case of statistical parametric TTS \cite{zen2013statistical}, it is a sequence of context vectors  $\boldsymbol{x}_{1:N}=(\boldsymbol{x}_{1}, \boldsymbol{x}_{2}, \cdots, \boldsymbol{x}_{N})$, where $N$ is the number of frames, and where the vector for the $n$-th frame is $\boldsymbol{x}_{n}$. Each $\boldsymbol{x}_{n}$ encodes linguistic information such as the phone and syllable identities. 
In the case of the MIDI-to-audio generation, the input MIDI contains messages that encode the time of note onset and offset, velocity, and other events. For processing using deep learning models, the MIDI input is usually represented as a piano roll\footnote{Although it is called piano roll, it can be used for other instruments.} that can also be written as a sequence of vectors $\boldsymbol{x}_{1:N}$, and each $\boldsymbol{x}_{n}$ is a  128-dimensional vector in which each dimension encodes the velocity for one of the 128 MIDI notes\footnote{The sustain pedal information can be reflected as elongation of notes, encoded by extending the note across multiple frames.}. Figure~\ref{fig:input_example} illustrates the difference.  

Accordingly, conversion from $\bs{x}_{1:N}$ to $\bs{o}_{1:T}$ can use similar models for both tasks. These methods can be grouped into two categories from the perspective of MIDI-to-audio generation. The first group is for applications where the audio is required to be consistent with the timing information in the MIDI input. In implementation, the synthesis process has to be $\boldsymbol{x}_{1:N} \rightarrow \boldsymbol{a}_{1:N} \rightarrow \boldsymbol{o}_{1:T}$, where the MIDI piano roll and acoustic feature sequences have the same length $N$, and where the audio waveform length $T$ is related to the feature length by a fixed frame shift $L$, i.e., $T=N\times{L}$. This pipeline is very similar to neural statistical parametric TTS  \cite{zen2013statistical}, and we can use various types of neural networks \cite{zen2014deep, ref:Fan14} for $\boldsymbol{x}_{1:N} \rightarrow \boldsymbol{a}_{1:N}$ and use neural waveform models \cite{Tamamori2017, prenger2018waveglow} or vocoders \cite{ref:Kawahara99, morise2016world} for $\boldsymbol{a}_{1:N} \rightarrow \boldsymbol{o}_{1:T}$. Note that $\boldsymbol{x}_{1:N}$ for TTS should have obtained the duration information from a duration model, while $\boldsymbol{x}_{1:N}$ for MIDI can retrieve the duration information from raw MIDI.

The second type of MIDI-to-audio application converts MIDI corresponding to a basic musical score into a professional ``performance'' with rich and dynamic expressivity, or transfers a particular ``performance style'' to the generated audio. In this case, $\boldsymbol{x}_{1:M}$ from input MIDI may not be aligned with the desired output sequence $\boldsymbol{a}_{1:N}$. Accordingly, the acoustic model should be capable of $\bs{x}_{1:M} \rightarrow \bs{a}_{1:N}$, and many attention-based sequence-to-sequence models \cite{Wang2017, li2019neural} are suitable for this task. 

Note that it is also possible to use a single model to directly convert the the input piano roll into an audio waveform, which is similar to WaveNet for TTS \cite{oord2016wavenet}.

\begin{figure}[t]
\centering
\includegraphics[trim=10 230 10 80, clip, width=0.46\textwidth]{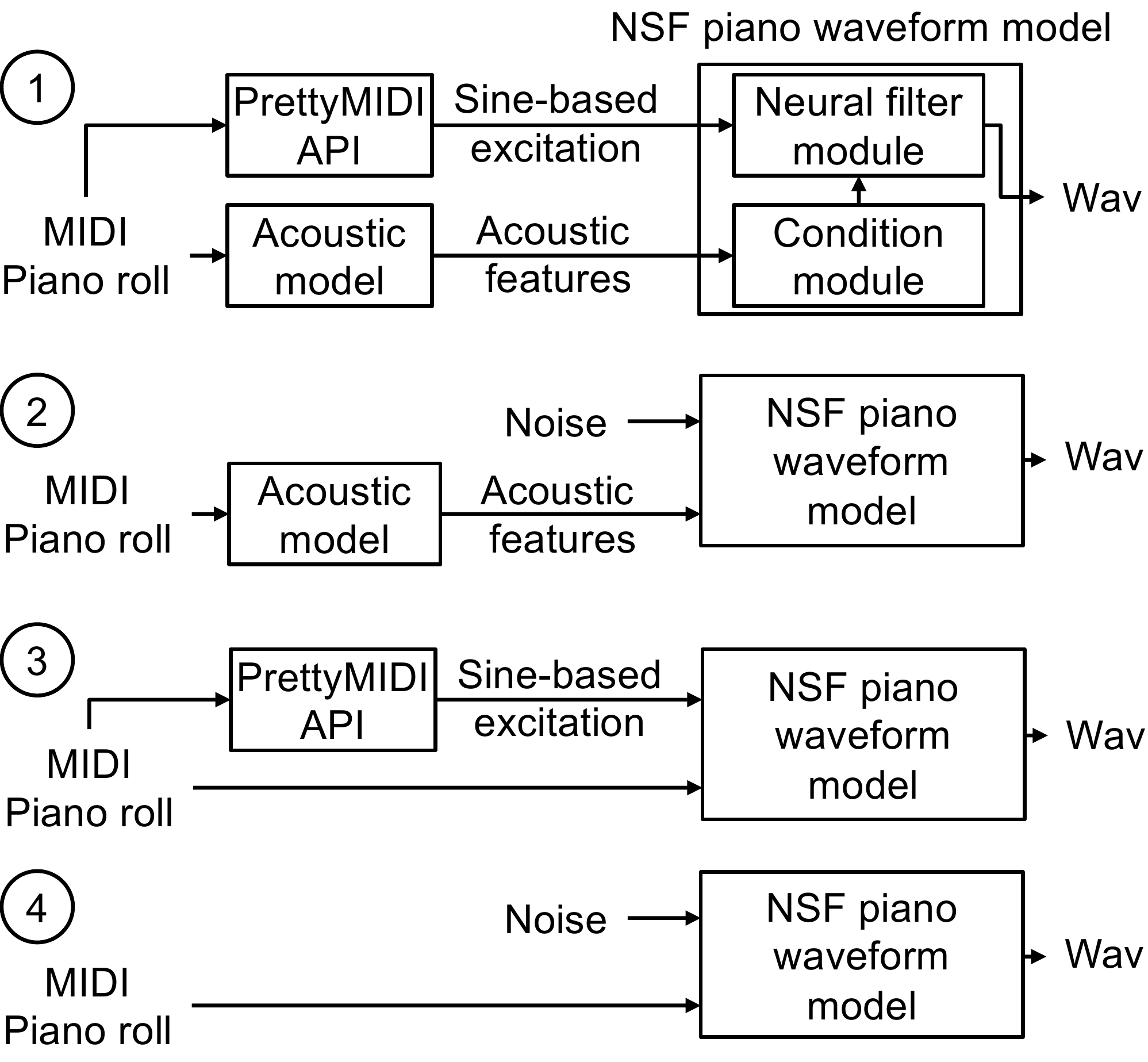}
\vspace{-3mm}
\caption{Four types of experimental MIDI-to-audio systems using NSF piano waveform model.}
\label{fig:model_example}
\vspace{-5mm}
\end{figure}

\begin{figure*}[t]
\centering
\includegraphics[width=0.96\textwidth]{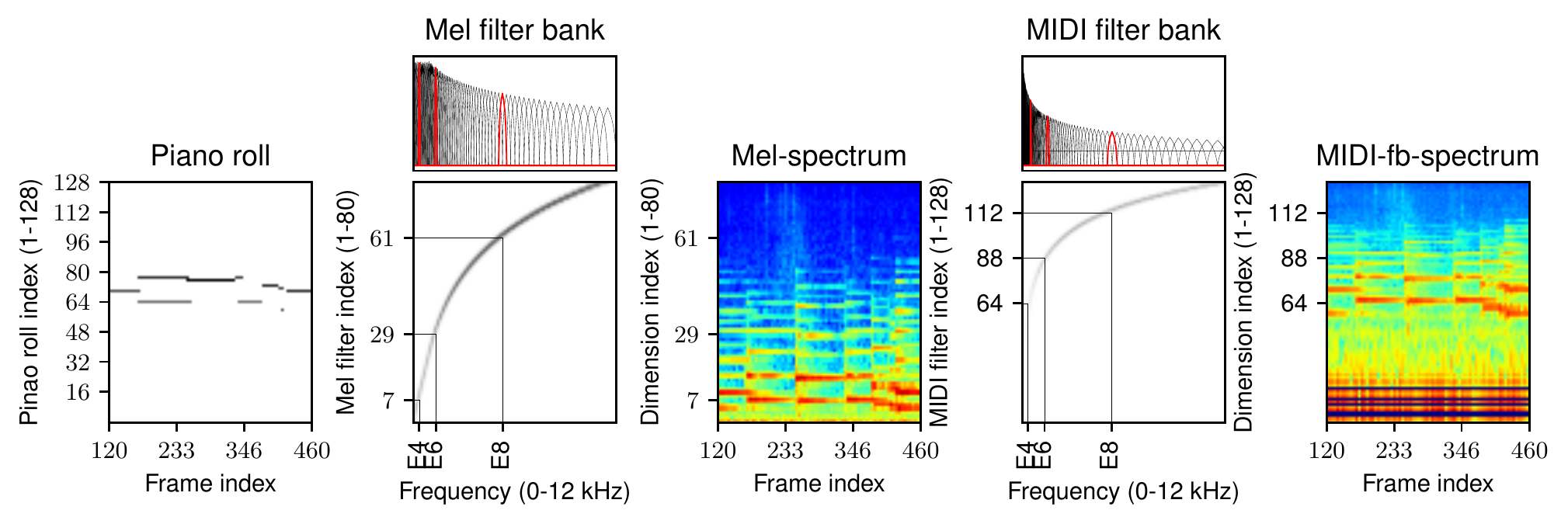}
\vspace{-5mm}
\caption{Mel-spectrogram versus MIDI-filter-bank spectrogram}
\label{fig:feat_example}
\vspace{-5mm}
\end{figure*}

\section{Experimental systems for time-aligned MIDI to piano audio synthesis}
\label{seq:3}
In this paper, we focus on systems that convert time-aligned MIDI to musical instrument audio waveforms and choose piano as the target instrument. Figure~\ref{fig:model_example} illustrates the systems as four groups. While all of them use an NSF-based waveform model, the first two types contain intermediate acoustic models, and the remaining two directly convert the input MIDI into an output waveform. Another difference among the systems is the type of the excitation signal, which will be detailed in Section~\ref{sec:nsf}. Note that the  experimental systems introduced in this section are applicable to  other monophonic or polyphonic instruments. 

\subsection{Acoustic models}
For the systems that use separate acoustic models, we can use many types of neural networks that conduct $\mathbb{R}^{N\times{D}}\rightarrow\mathbb{R}^{N\times{D'}}$. Although it is originally designed for sequence-to-sequence mapping, we investigate variants of Tacotron, hoping that the outcome can accelerate our research on the unaligned MIDI-to-audio synthesis task in the near future.

Our Tacotron acoustic model implementation was based on  \cite{yasuda2019investigation}.  Tacotron was modified to accept a sequence of 128-dimensional MIDI piano roll frames as input instead of a text or phoneme sequence.  Tacotron's encoder learns an embedding that maps a symbolic token to an embedding vector, but since piano roll frames are not strictly symbolic but are already a meaningful vector representation of pitch and velocity, we replaced this embedding layer with a dense projection layer.  Aside from these initial modifications of the encoder to accept MIDI piano roll input instead of text, the model architecture is otherwise the same as in \cite{yasuda2019investigation}. The encoder consists of a CBHG module followed by a self-attention block.  The outputs of both are input to the decoder, where the CBHG output is input to a forward attention mechanism and the self-attention output is received by an additive attention mechanism.  The decoder consists of a decoder recurrent layer followed by self-attention, and finally, a post-net conducts spectral shaping and enhancement for the final spectrogram output.

Because music sequences are much longer than the typical short utterances used for training text-to-speech synthesis models, and in particular, the piano roll input representation is the same length as the output spectrogram, as opposed to text or phoneme sequences which are much shorter, we had to take a number of steps to fit the training sequences into memory and to learn alignments well.  First, we found that it was necessary to segment the data into shorter sequences, starting with 200 frames.  Next, we wished to reduce the autoregressive dependencies and instead force the model to rely more directly on the inputs.  The prenet to the decoder receives the previous predicted spectrogram frame (or the previous actual spectrogram frame, when teacher-forcing is enabled), so we increased dropout at the prenet to the decoder from its default value of 0.5 to higher values of 0.9, 0.95, or 0.99, finding in initial experiments that the 0.99 dropout rate produced the best-aligned predictions.  Next, we tried a downsampling approach where we downsampled the input piano roll sequence by a factor of either 2 or 4, and effectively downsampled the output spectrogram as well by setting the reduction factor to 2 or 4, so that the model predicts that many output frames at each timestep.  We found that a downsampling factor of 4 produced the best alignments, and that combining this downsampling with the dropout to the decoder prenet at a rate of 0.99 produced stable alignments for sequences as long as 800 frames, so we chose this as our base model, which we call \texttt{taco2}.  

As an additional method to force the model to use the input sequence more directly, we concatenated the current MIDI piano roll frame with the previous spectrogram frame at the decoder prenet (after that spectrogram frame has been dropped out).  We call this model configuration \texttt{taco3}, and we warm-start its parameters from \texttt{taco2}.  Finally, to see whether up-sampling the data again would improve the synthesis quality, model \texttt{taco4} uses none of these prenet or downsampling tricks, and its parameters are also warm-started from \texttt{taco2}.

As reference, we included a CNN-based network that conducts  $\mathbb{R}^{N\times{D}}\rightarrow\mathbb{R}^{N\times{D'}}$. This network called PerformanceNet combines U-Net and multi-band convolution blocks to convert MIDI piano rolls into acoustic features and has shown good performance on MIDI-to-audio synthesis \cite{wang2019performancenet}.  Samples from all systems can be heard online\footnote{https://nii-yamagishilab.github.io/samples-xin/main-midi2audio.html}.

\subsection{MIDI filter-bank features}
\label{seq:midifb}
Many TTS systems use Mel-spectrogram or Mel-cepstral coefficients as the output of the acoustic model, but the Mel scale may not be the best for music applications. 
Since each MIDI note $d$ and its corresponding frequency $f$ is related by
\begin{equation}
    f = 2^{\frac{d-69}{12}}\times440,
\end{equation}
where 69 and 440 are the MIDI index and frequency value of note A4, respectively,
we can define a new filter bank where the $k$-th filter is centered around $f=2^{(k-69)/12}\times440$ Hz. 
Figure~\ref{fig:feat_example} plots the resulting MIDI-based filter bank.

We apply the MIDI-based filter bank to extract low-dimensional spectral features in a similar manner to the Mel-spectrogram.
Figure~\ref{fig:feat_example} also compares the Mel-spectrogram and MIDI-filter-bank-based spectra. Notice that the bars in the MIDI-filter-bank-based spectra resemble the corresponding piano roll.
Note that the MIDI-filter-bank spectra have empty dimensions in the low frequency region because some of the filters do not cover any discrete frequency bin. These empty dimensions can be filled in if we increase the FFT points.

\begin{figure}[t]
\centering
{\includegraphics[trim=10 550 10 75, clip, width=0.48\textwidth]{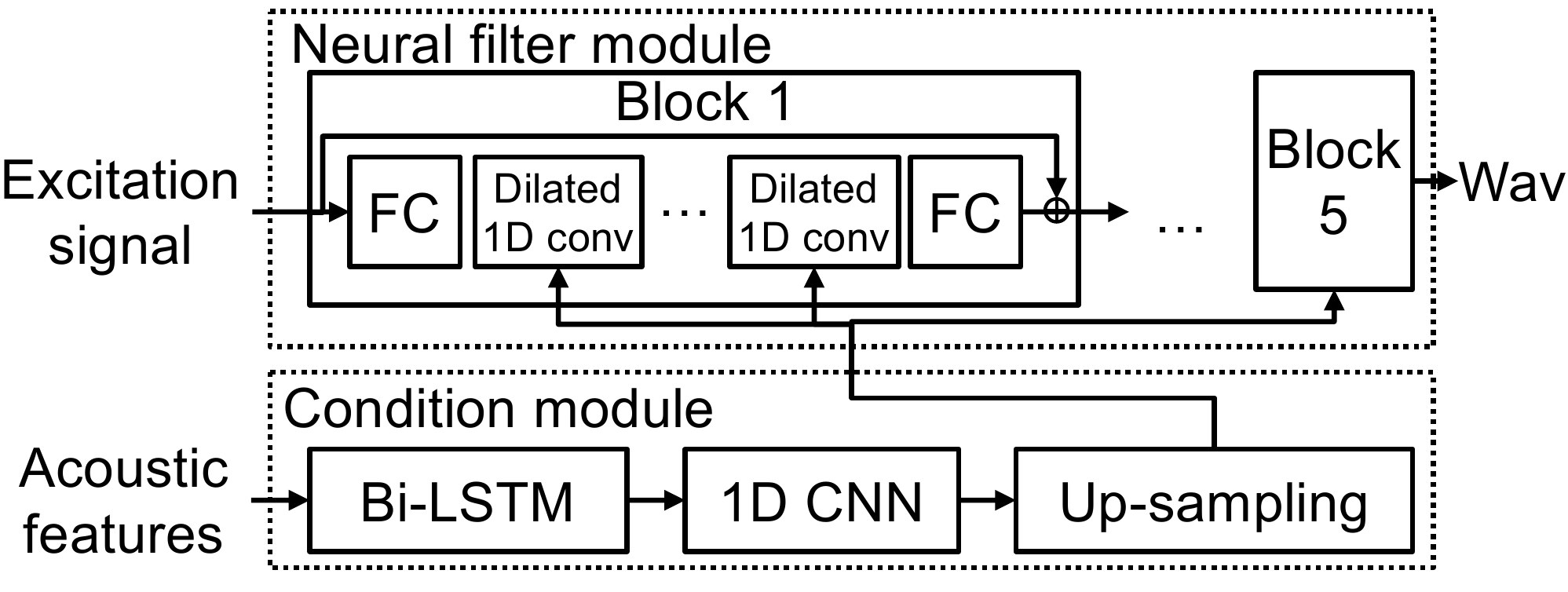}}
\vspace{-5mm}
\caption{NSF piano waveform model. Block in neural filter module is based on the simplified dilated-CNN filter block (cf. Figure.4 in \cite{wangNSFall}). FC denotes a fully-connected layer.}
\label{fig:nsf_example}
\vspace{-5mm}
\end{figure}

\subsection{NSF-based piano waveform model}
\label{sec:nsf}
Our NSF-based piano waveform model is based on the simplified NSF model \cite{wangNSFall}. As Fig.~\ref{fig:nsf_example} illustrates, the NSF piano waveform model contains a condition module that transforms and up-samples the input frame-level acoustic features and a neural filter module that converts the up-sampled features and an excitation signal into an output waveform through multiple dilated convolution blocks. 

A major difference between NSF waveform models for piano and speech is the source module. While the source module for speech produces an excitation signal from fundamental frequencies (F0s), such a module cannot be used for polyphonic piano sound. One solution is to use noise as the excitation (e.g., \textcirclednew{2} and \textcirclednew{4} in Fig.~\ref{fig:model_example}).
An alternative solution is to derive a sine-based excitation signal from the input time-aligned MIDI. In this paper, we use the PrettyMIDI synthesis API \cite{raffel2014intuitive}\footnote{https://craffel.github.io/pretty-midi/} to produce a polyphonic sine-based excitation signal (e.g., \textcirclednew{1} and \textcirclednew{3} in Fig.~\ref{fig:model_example}) from the piano roll notes.

\begin{figure*}[t]
\centering
\includegraphics[width=0.96\textwidth]{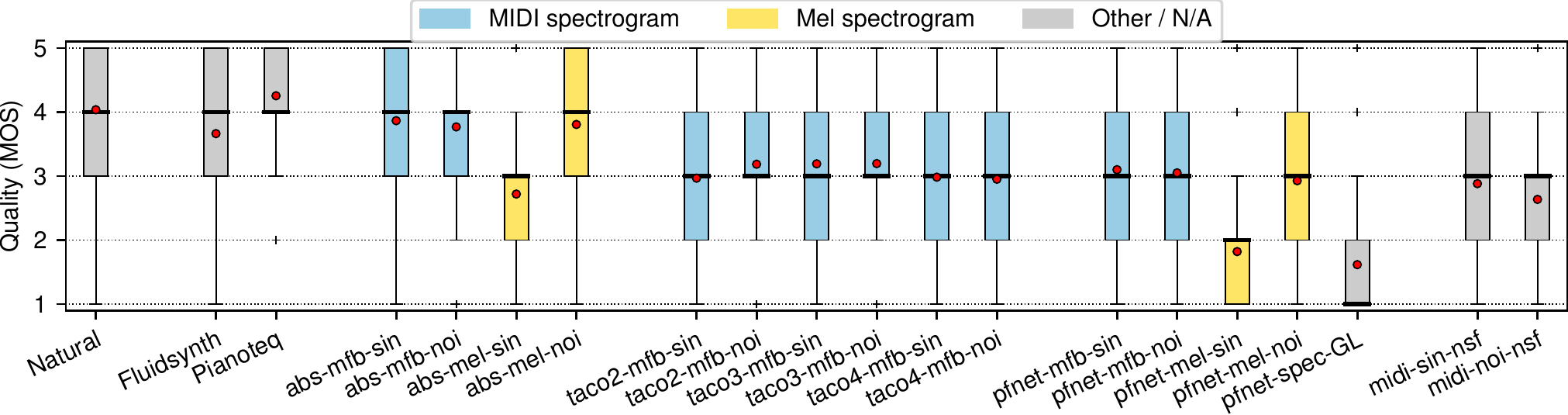}
\vspace{-3mm}
\caption{Boxplots of MOS per system. Red dots denote mean of MOS.}
\label{fig:mos_test}
\vspace{-5mm}
\end{figure*}

\section{Experiments}
\subsection{Database and protocol}
Experiments were conducted using the MAESTRO database (V2.0.0)\footnote{https://magenta.tensorflow.org/datasets/maestro} \cite{hawthorne2018enabling}. This is a large-scale database that contains over 200 hours of piano performances and aligned MIDI data from the International Piano-e-Competition.
Both the audio and MIDI data were recorded when the competing virtuoso pianists performed on concert-quality acoustic grand pianos with integrated MIDI capture and playback systems. 

The experiments followed the official data protocol: a train set with 161.3 hours of data from 967 performances, a validation set with 19.4 hours of data from 137 performances, and a test set with 20.5 hours of data. Because it is impossible to evaluate the entire test set in subjective evaluation, 192 test segments were manually excerpted from the test set, and each test segment was less than 30 seconds in duration. 

\subsection{Experimental systems and feature configurations}
Table~\ref{tab:models} lists the systems investigated in the experiments. The first two are reference software synthesizers, and the next four are copy-synthesis systems that directly use natural acoustic features (i.e., Mel-spectrogram or MIDI-based filter bank features) as the input to the NSF model for piano waveform generation. They simulate the ideal case where acoustic models convert the input MIDI into the acoustic features perfectly. The next 11 systems are pipelines of an acoustic model, which is either a variant of the Tacotron or the PerformanceNet model, and the NSF waveform model. The last two experimental systems, namely \sysII\ and \sysI, directly convert the MIDI and the excitation signals into the waveform through NSF models.

We trained Tacotron models using the MIDI filter bank spectrogram as output, since we found initially that this produced better alignments than using Mel spectrograms.\footnote{Alignment errors were reduced from 21\% using Mel spectrograms to 3\% using MIDI spectrograms.}  The models were trained on segments of 800 frames using the Adam optimizer, a batch size of 4, and a learning rate of 0.0001.  The base model \texttt{taco2} was trained for 550k steps until spectrogram loss on the development set had converged.  The other two models \texttt{taco3} and \texttt{taco4} had their parameters warm-started from \texttt{taco2}, and were trained for an additional 250k steps and 50k steps to convergence, respectively.
The PerformanceNet-based acoustic models (\texttt{PFNet}) were trained for 50 epochs using the Adam optimizer with a batch size of 16.

All the NSF models were trained using an Adam optimizer with a learning rate of $1\times10^{-4}$ and \cite{kingma2014adam}. 
The maximum number of training epochs is 20, and each epoch took around 24 hours.
Due to the limited GPU memory size, the input and output data for NSF models was truncated into segments of 3s in duration, and the batch size was set to 5. During generation, the NSF models produced the waveform without truncation.

As a reference, the original PerformanceNet was included in the experiment. This system uses spectrograms as the acoustic feature and produces a waveform from the generated spectrogram through the Griffin-Lim (GL) algorithm \cite{griffin1984signal}.
Two software synthesizers were also included for reference: an open-source software called \texttt{Fluidsynth}\footnote{https://www.fluidsynth.org/} and a commercial one called \sysXV\footnote{https://www.modartt.com/pianoteq}. While both produce piano audio given MIDI input, the former uses a sampling-based approach, and the latter is based on physical modeling of pianos.

\begin{table}[t]
\centering
\caption{Experimental systems. Pitch accuracy is measured using cross-entropy, the lower the better. }
\vspace{-2mm}
\label{tab:models}
\resizebox{\columnwidth}{!}{
\setlength{\tabcolsep}{2pt}
\begin{tabular}{cccccccc}
\toprule
\multirow{2}{*}{System ID} & \multirow{2}{*}{\shortstack{Acoustic \\ model}} & \multirow{2}{*}{\shortstack{Acoustic \\ feature}} & \multirow{2}{*}{\shortstack{Excit. \\ signal}} & \multirow{2}{*}{\shortstack{Wave. \\ model}} & \multicolumn{2}{c}{Pitch mismatch} & \multirow{2}{*}{\shortstack{MOS \\ (mean)}}\\
\cline{6-7}
                    & & &  &   & note & chord & \\
\midrule
Natural & - & - & - & - & - & - & 4.04 \\
\midrule
\sysXIV & \multicolumn{4}{l}{Sample-based MIDI-to-audio software} &  5.20 &  6.77  & 3.66 \\
\sysXV & \multicolumn{4}{l}{Physical-model MIDI-to-audio software} &  4.82 &  6.50 & {4.25} \\
\midrule
\absIV & - & midi-fb & sine & NSF   & - & -   & 3.87 \\
\absII & - & midi-fb & noise & NSF  & - & -   & 3.77 \\
\absIII & - & mel-spc. & sine & NSF & - & -   & 2.72 \\
\absI & - & mel-spc. & noise & NSF  & - & -   & 3.81 \\
\midrule
\sysVIII & \texttt{taco2} & midi-fb & sine & NSF 	&  {4.61} &  6.34 & 2.97 \\
\sysVII & \texttt{taco2} & midi-fb & noise & NSF 	&  4.66 &  6.36 & 3.18 \\
\sysX  & \texttt{taco3} & midi-fb & sine & NSF 		&  4.78 &  6.48 & 3.19 \\
\sysIX & \texttt{taco3} & midi-fb & noise & NSF 	&  4.89 &  6.53 & 3.19 \\
\sysXII & \texttt{taco4} & midi-fb & sine & NSF 	&  4.86 &  6.39 & 2.98 \\
\sysXI  & \texttt{taco4} & midi-fb & noise & NSF 	&  4.97 &  6.42 & 2.95 \\
\sysVI & PFNet & midi-fb & sine & NSF 				&  5.59 &  7.14 & 3.10 \\
\sysIV & PFNet & midi-fb & noise & NSF 				&  5.78 &  7.26 & 3.05 \\
\sysV & PFNet & mel-spec. & sine & NSF 				&  5.66 &  7.17 & 1.82 \\
\sysIII & PFNet & mel-spec. & noise & NSF 			&  5.74 &  7.25 & 2.93 \\
\sysXIII  & PFNet & spec. & - & GL 					& 5.43    & 6.98			& 1.62 \\
\midrule
\sysII & - & - & sine & NSF &  4.32 &  6.40  & 2.88 \\
\sysI & - & - & noise & NSF  &  4.40 &  6.08 & 2.63 \\
\bottomrule
\end{tabular}}
\vspace{-5mm}
\end{table}

The audio waveforms from MAESTRO were down-sampled to 24kHz and encoded through 16-bit PCM. The Mel-spectrogram was then extracted using a frame length of 50ms, a frame shift of 12 ms, FFT with 2048 points, and 80 overlapped triangular filters evenly spaced on the Mel-frequency scale. The MIDI-based filter bank features were extracted using a similar configuration but with a filter bank based on the MIDI notes (Section~\ref{seq:midifb}). The spectrogram for \sysXIII\ used the original recipe \cite{wang2019performancenet}. The MIDI files were converted into 128-dimensional piano rolls using the PrettyMIDI API.


\begin{figure}[t]
\centering
\includegraphics[width=0.4\textwidth]{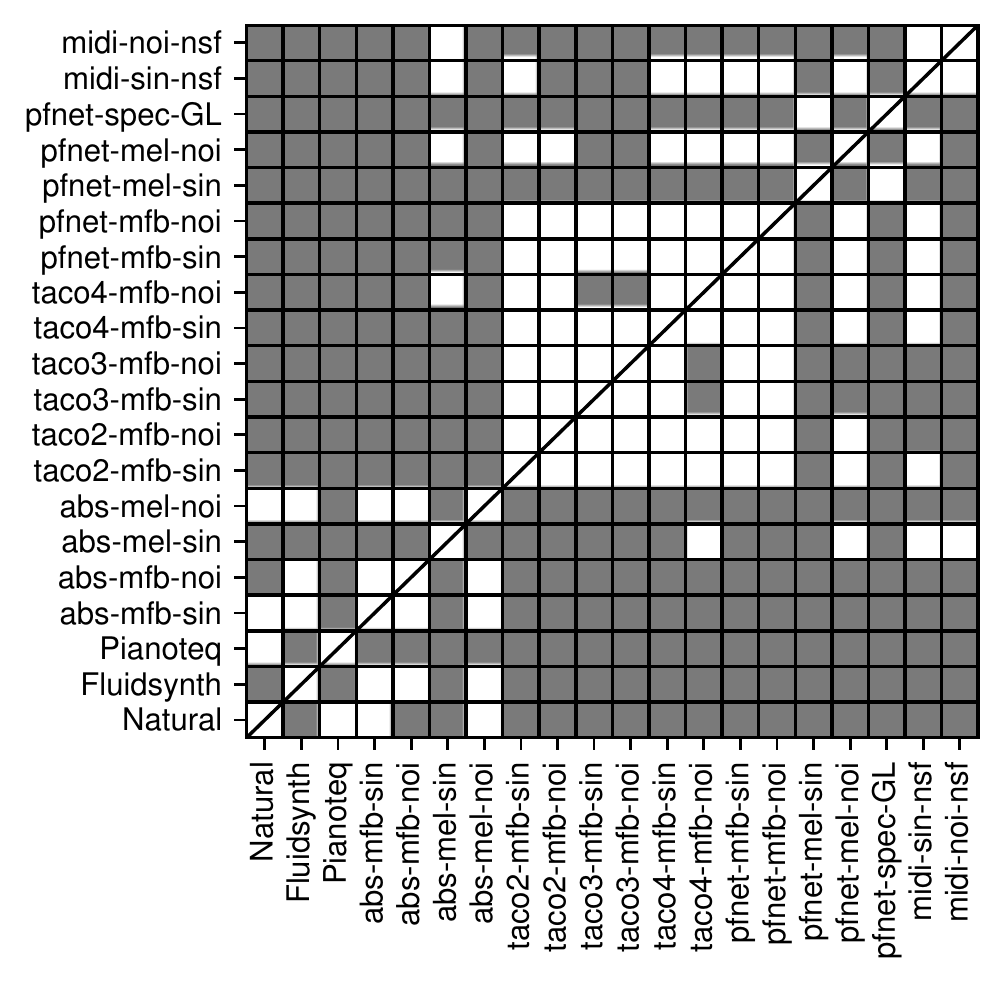}
\vspace{-3mm}
\caption{Results of two-sided Mann-Whitney test with Holm-Bonferroni correction. A grey block indicates a statistically significant difference at $\alpha=0.05$. }
\label{fig:sig_test}
\vspace{-5mm}
\end{figure}

\subsection{Subjective evaluation and results}

We conducted a crowdsourced listening test to evaluate the quality of audio from our synthesizers, comparison synthesizers, and natural audio.  We included 120 samples from each of the 20 systems, and obtained mean opinion score (MOS) ratings for each sample from five different listeners on a scale from 1 (very bad) to 5 (very good).  Listeners were instructed to evaluate the overall quality of the piano sound subjectively.  Each test set contained 30 different samples, balanced to contain at least one sample from each of the 20 systems. Listeners were permitted to complete up to 10 different sets.  In total, we received ratings from 224 unique listeners.  A box plot of the MOS ratings for each of the 20 systems can be seen in Figure \ref{fig:mos_test}, and significant differences from a Holm-Bonferroni-corrected two-sided Mann-Whitney U test at a level of $\alpha=0.05$ are shown in Figure \ref{fig:sig_test}. The mean of MOS is also listed in Table~\ref{tab:models}.

We found that the \sysXV\ physical synthesis was significantly preferred over every other synthesis method, and was even rated higher than natural piano audio (although not significantly so).  Other systems which were not significantly different from natural audio were two of the analysis-by-synthesis systems, \absIV\ and \absI.  There were not many statistically-significant differences between the Tacotron models, with the exception of \sysXI, which was significantly worse than both of the \texttt{taco3} systems.  As for PFNet-based systems, the two that used the MIDI filter bank representation had about equivalent performance to the Tacotrons with no significant differences, whereas \sysV\ and \sysXIII\  were significantly worse than all Tacotrons.

Comparing the standard Mel filter bank to the proposed MIDI filter bank, there are two significant differences favoring MIDI, \sysV\ vs.~\sysVI\ and \absIII\ vs.~\absIV.  For noise vs.~sine wave excitation, there are two significant differences favoring noise, \sysV\ vs.~\sysIII\ and \absIII\ vs.~\absI.

\subsection{Objective evaluation and results}
We first measured the pitch accuracy of the synthesized audio. We trained a CNN-based F0 estimator called CREPE \cite{kim2018crepe} on the MAESTRO training set. Although the original CREPE is designed for monophonic sound, it can be modified for polyphonic piano sound by replacing the target from one-hot vectors to  multi-hot ones. During the pitch detection stage, we extract the pitch probability sequence $\bs{p}_{1:N}=(\bs{p}_1,\cdots,\bs{p}_N)$ from the input audio $\bs{o}_{1:T}$ by $\bs{p}_{1:N}=\texttt{CREPE}(\bs{o}_{1:T})$, where each $\bs{p}_n=[{p}_{n,1},\cdots,{p}_{n,128}]$, and where ${p}_{n,k}\in(0,1)$ indicates the probability of observing the $k$-th MIDI note at the $n$-th frame. Then, the cross entropy between $\bs{p}_{1:N}$ and the input piano roll $\bs{x}_{1:N}$ can be computed to measure the pitch mismatch $\texttt{CE} = -\sum_{n=1}^{N}\sum_{k=1}^{128}x_{n,k}\log p_{n,k}$. Hence, a lower $\texttt{CE}$ indicate less severe mismatch.

We created piano rolls and synthesized audio for around 100 individual notes and chords and synthesized their audio. We measured cross entropy and listed results in Table~\ref{tab:models}. 
It can be observed that, when using separate acoustic and waveform models, the systems with sine excitation outperformed their counterparts using noise excitation, but the systems using only the NSF waveform model achieved lower $\texttt{CE}$ values. Furthermore, Tacotron models have lower mismatches compared to PFNet systems using the same vocoder.  However, lower pitch mismatch does not lead to higher MOS. This indicates that the perceptual quality of piano audio is not only affected by pitch accuracy. Another hypothesis is that amateur listeners may not be able to detect mild pitch mismatch.

\section{Discussion}
The evaluation results suggest that the physical-model-based approach outperformed the other deep-learning-based MIDI-to-audio systems and is even slightly better than the original audio in MAESTRO. The original audio was recorded over many years, and we observed that the MOS of the audio in year 2008 and 2014 were less than 4.0. This variation of quality may also affect the MAESTRO training set and the deep-learning-based models trained using this data. On the other hand, the physical-model-based approach is free from such artifacts in data.  However, the physical model is the outcome of laborious analysis and simulation \cite{chaigne1994numerical, Giordano2004, chabassier2012physical}, which does not easily generalize to another piano type or instrument. In contrast, deep-learning-based models are more flexible, and this study showed examples of using TTS models for music generation. 

Performance of the investigated deep-learning models is not satisfying yet. Since listeners may be more sensitive to the artifacts in music signals, this sets a high standard for producing natural audio but also leaves large room for improvement. One potential direction for future work is to combine data-driven techniques with physical models of piano sounds.  Rather than learning a physical sound from scratch, sound physics may offer effective prior knowledge. 

\section{Conclusions}
This study is our initial step to investigate the possibility of using TTS models for MIDI-to-audio synthesis. The two disciplines differ in many aspects but both deal with the mapping from one sequence of data into another. Based on the similarities and differences, we modified the TTS components, namely Tacotron and NSF, and introduced a MIDI filter-bank acoustic feature set for the MIDI-to-audio task, which improved alignments for Tacotron and resulted in more preferred audio. Based on subjective evaluation of the TTS-like systems, natural audio, physical piano model, and other reference systems, we observed promising results when using TTS components for MIDI-to-audio generation. We also identified the quality bottleneck when converting the MIDI input into acoustic features, a similar bottleneck to that in TTS. Hence, future work will explore more different types of acoustic and waveform models, and we encourage TTS researchers to extend their knowledge and practices to this challenging task of MIDI-to-audio generation.

%
%
%
%
%
%
%
%

\section{Acknowledgements}
A part of the computations were performed on the TSUBAME 3.0 supercomputer at Tokyo Institute of Technology.  This work is supported by “TSUBAME Encouragement Program for Young/Female Users” of Global Scientific Information and Computing Center at Tokyo Institute of Technology and by “Joint Usage/Research Center for Interdisciplinary Large-scale Information Infrastructures” in Japan, and by JST CREST Grants (JPMJCR18A6 and JPMJCR20D3),  JST AIP Challenge Grant, and MEXT KAKENHI Grants (18H04112, 21H04906, 21K11951, 21K17775), and a grant by KAWAI foundation for sound technology \& music (year 2020, No.4), Japan.


\bibliographystyle{IEEEtran}

\bibliography{mybib}


\end{document}